\documentclass[a4paper,12pt]{article}
\usepackage{cite}
\usepackage{epsfig, amsmath, amssymb}
\usepackage{graphicx,psfrag}
\usepackage{xcolor} 
\usepackage{graphicx}
\usepackage{amssymb}
\usepackage{amsmath}
\usepackage{amsfonts}
\usepackage{dsfont}
\usepackage{mathtools}
\usepackage{slashed}
\usepackage{amsmath}
\usepackage{caption}
\usepackage{tikz}
\usepackage{rotating}
\usepackage{bbold,amsfonts}

\usepackage[utf8]{inputenc}
\usepackage{bm}
\usepackage{xcolor}
\usepackage{float}
\usepackage{braket}
\numberwithin{equation}{section}
\usepackage[height=8.8in,width=6.45in]{geometry}
\usepackage[font=small,labelfont=bf]{caption}
\usepackage[hidelinks]{hyperref}
\usepackage{mathtools}

\newcommand\vertarrowbox[3][6ex]{%
	\begin{array}[t]{@{}c@{}} #2 \\
		\left\downarrow\vcenter{\hrule height #1}\right.\kern-\nulldelimiterspace\\
		\makebox[0pt]{\scriptsize#3}
	\end{array}%
}
\begin{document}
	\begin{titlepage}
	\vbox{
		\halign{#\hfil         \cr
		} 
	}  
	\vspace*{15mm}
	\begin{center}
		{\Large \bf 
			More Holographic M5 branes in $AdS_7 \times S^4$
		}
		
		\vspace*{15mm}
		
		{\large Varun Gupta}
		\vspace*{8mm}
		
		Chennai Mathematical Institute, \\
		SIPCOT IT Park, Siruseri 603103, India \\
		
		\vskip 0.8cm
		
		
		{\small
			E-mail:  vgupta@cmi.ac.in
		}
		\vspace*{0.8cm}
	\end{center}
	
	\begin{abstract}
		
	We study classical M5 brane solutions in the probe limit in the $AdS_7 \times S^4$ spacetime geometry with worldvolume 3-form flux. These solutions describe the holography of codimension-4 defects in the 6d boundary dual $\mathcal{N}= (0,2)$ supersymmetric gauge theories. Starting from some half-BPS solutions which follow stricter BPS conditions, we find a general $\frac1{16}$-BPS solution. We show how some of the \textit{giant-like} M5 brane solutions here may be related to the codimension-2 surface operators in the 4d gauge theory.
		

	\end{abstract}
	\vskip 1cm
	{
	}
\end{titlepage}

	 	\setcounter{tocdepth}{2}
	 \tableofcontents
	 \vspace{.5cm}
	 \begingroup
	 \allowdisplaybreaks
	 
	 \section{Introduction}
	 The study of branes as probes in the anti-de Sitter spacetime provides valuable information about the non-local observables such as surface operators, Wilson lines, and other line operators in the boundary supersymmetric gauge theory \cite{Drukker:2005kx, Drukker:2008wr, Koh:2008kt}, \cite{Howe:1997ue, Maldacena:98}. In the M-theory framework, using the holographic  $AdS_7 / CFT_6$ correspondence, M5 branes have been used as probes to know more about the non-local Wilson surface operators \cite{Lunin:2007, Chen:2007} which are codimension-4 in the 6d boundary gauge theory, and as well as the codimension-2 defects in the theory\cite{FGT:2015, VG:21, Gutperle:22}. Codimension-4 Wilson surface operators have also been analyzed due to the holography of probe M2 branes \cite{Drukkeretal:0320, Drukkeretal:0420}. The codimension-2 defects upon the dimensional reduction of the theory on a Torus $\mathbb{T}^2$ become Gukov-Witten surface defects in the 4d gauge theory \cite{Gukov:2006jk, Gukov:2008sn}, \cite{GGS:2013, Gukov:2014, FGT:2015, Leflochetal:0217, SKAMBetal:0717, SKAMBetal:0718}. Whereas, in the literature, it has been discussed that the codimension-4 defects upon a suitable dimensional reduction may become Wilson lines in the large rank representations of the 5d SYM gauge theory \cite{MoriYamaguchi:2014}, or they may become the codimension-2 surface operators in the 4d gauge theory \cite{FGT:2015, Gukov:2014}.
	 In \cite{FGT:2015} authors show how surface operators in the 4d gauge theory obtained from the codimension-2 and codimension-4 non-local operators are equivalent objects by showing the equivalence of the twisted-chiral superpotential functions that describe them in the IR regime. \\
	 
	  In \cite{VG:21}, we analyzed those M5 brane embedding solutions that are holographic duals of the codimension-2 surface defects and they end in the boundary of the global $AdS_7$ in an $\mathbb{R} \times \mathbb{S}^3$ submanifold. We found the most general BPS M5 brane solutions that preserved just one spacetime supersymmetry. Out of these general $\frac1{32}$-BPS solutions, there were some solutions with at least 2 preserved supercharges that were shown to hit the $AdS_7$ boundary in $1$ $+$ $3$ submanifold. For these $\frac1{16}$-BPS solutions, we also recovered the singularity profile of the complex scalar in the $(2,0)$ tensor multiplet theory when we took the large radius limit for the $AdS_7$ boundary. \\
	  
	  In this article, the point of discussion is towards the codimension-4 surface operators in the 6d gauge theory. The holographic dual M5 branes hit the boundary of $AdS_7$ in a $1\, +\, 1$ dimensional submanifold. The M5 brane solutions that we analyze here in the probe limit can have self-dual 3-form flux turned on their world volume. Half-BPS solutions have the world volume of $AdS_3 \times S^3$, where the flux strength is proportional to the $S^3$ volume. There are two kinds of solutions possible here: the first kinds are the ones for which the $S^3$ part of the worldvolume is inscribed in the $S^4$ part of the background global $AdS_7 \times S^4$ spacetime geometry; the second kinds are the ones for which all of the $AdS_3 \times S^3$ worldvolume coordinates are identified with the $AdS_7$ coordinates. We will often refer to the first ones as the ``giant-like" solutions and the second ones as the ``dual-giant-like". \\
	  
	  The ``giant-like" solutions that we find here are similar to the ones discussed in \cite{Chen:2007, Lunin:2007}. And like the ``giant-like" solutions known from earlier, we are finding that the 3-form flux can be turned off to zero value for a certain instance, and the M5 world volume is still a stable supersymmetric solution.
	 Another important result here is that we show these ``giant-like" solutions are part of the most general $\frac1{32}$-BPS embedding conditions found in \cite{VG:21} when we analyzed the M5 brane solutions, which were the duals of codimension-2 defects in the boundary theory. We will give more comments on the relation with our previous results in subsection \ref{common susy} and in the conclusion section \ref{conclusionsection}. The ``dual-giant-like" solutions that we find are also similar to the solutions that authors in \cite{Chen:2007} have found, and over there, the presence of the 3-form flux is indeed necessary to have a supersymmetric world volume. In section 2, we begin by describing the coordinate frame we choose for the $AdS_7\times S^4$ metric to find the desired solutions. In section 3, we give the details about the ``giant-like" M5 solutions that we find and then show the possibility of combining them in ways so that their \textit{intersections} preserve the common set of supercharges. The general ``giant-like" M5 solutions preserve just 2 common supercharges which we derive in subsection \ref{common susy} and App \ref{AppA} and present in the equation \eqref{newgensolution}. In section 4, we give the details of the ``dual-giant-like" solutions that we find. In section 5, we end with an analysis of our results with some extra details and future outlooks from this work.
	\section{Probe M5s in $AdS_7 \times S^4$ geometry}
		
		\noindent
		The metric that we consider for the eleven-dimensional $AdS_7 \times S^4$ geometry is in the global coordinates system
		\begin{align}
		\label{gddbulk}
		ds^2_{AdS} = - \left(1 + \frac{r^2}{4l^2}  \right) dt^2 + \frac{dr^2}{\left(1 + \frac{r^2}{4l^2} \right) } + r^2 d\Omega_5
		\end{align}
		with $d\Omega_5 = d\alpha^2 + \cos^2 \alpha \, d\phi_1^2  + \sin^2 \alpha \left( d\beta^2 + \cos^2 \beta \, d\phi_2^2 + \sin^2 \beta \, d\phi_3^2 \right) $
		\begin{align}
		\label{S4metric}
		ds^2_{S^4} =  l^2 \left( d\theta^2 +  \sin^2 \theta ( d\chi^2 + \cos^2 \chi \, d\xi_1^2  + \sin^2 \chi \, d\xi_2^2 ) \right)
		\end{align}
		%
		%
		%
		The background 3-form potential field is given by
		\begin{align}
		C^{(3)} = \frac{l^3}2 \cos \theta \left( \cos 2 \theta \,-\, 5 \right)\, d\Omega_3
		\end{align}
		%
		%
		
	\subsection*{Half-BPS M5 branes}\label{halfBPS}
We now spell out the type of half-BPS M5 brane embeddings that we are going to analyze. Such solutions will have the worldvolume of $AdS_3 \times S^3$ metric. To find the desired supersymmetric M5 solutions, we follow the method employed in \cite{Chen:2007} which followed the classical equations obtained in the work \cite{Howeetal:1997} where the worldvolume superspace was considered to be embedded in the 11-dimensional target superspace. These field equations have been shown to be the same as Euler-Lagrange equations of motion \cite{PSTetal:1997},  from the covariant action of Pasti-Sorokin-Tonin (PST) \cite{PST:1997, PSTetal:1997pt1}, which is a Dirac-Born-Infeld like covariant action. Here we only present the analysis of the $\kappa$-symmetry constraint equation but all the half-BPS solutions to be discussed solve all the remaining classical equations. The $S^3$ part of the induced metric has been understood to support the 3-form flux of the worldvolume.
We will consider two kinds of solutions: \\

\noindent
 i) \textit{giant-like}: the ones which have the $S^3$ sphere expressed in terms of the coordinates that parametrize $S^4$ of the 11d spacetime. \\
 
\noindent
ii) \textit{dual-giant-like}: the second kinds are the ones where all the M5 coordinates comes from $\subset$ $AdS_7$ $coordinates$ of 11d. \\

The three-form flux here is self-dual and follows the constraint $h$ $= \star_g h$ \footnote{Flux field $h$ here is related to the gauge-invariant field strength $H$ as equation (28) of reference \cite{PSTetal:1997}. And $H 
 = C^{(3)} \, + \, d B^{(2)}$ where $B^{(2)}$ is the gauge potential that describes the three on-shell degrees of freedom of the M5 worldvolume theory.}. Hence we can derive a form for the field strength desired for our analysis throughout this note
\begin{align}
	h = h_{\tau12} d\tau \wedge d\sigma_1 \wedge d\sigma_2 \, + \, h_{345} d\sigma_3 \wedge d\sigma_4 \wedge d\sigma_5
\end{align}
Worldvolume self-duality constraint allows us to write this in terms of the explicit expression
\begin{align}
\label{genselfdh}
h = a \, \left( - \frac{\sqrt{- \text{det} \, g}}{ \sqrt{\Omega_3} } \, d\tau \wedge d\sigma_1 \wedge d\sigma_2 \, + \, \sqrt{ \Omega_3} \, d\sigma_3 \wedge d\sigma_4 \wedge d\sigma_5 \right)
\end{align}
where $\Omega_3$ is the volume of the $S^3$ and the notation $g$ is for the induced worldvolume metric. $a$ is some common factor that turns out to be a constant in both the kinds of solutions that we discuss here. \\

\noindent
We will begin by enumerating all the possible solutions that we found in the first category.
\vspace{.3cm}
\section{Solutions of the kind I :  $S^3$ $\subset$ $S^4$}\label{giant-like}

The \textit{giant-like} M5 solutions which appeared in the literature \cite{Lunin:2007, Chen:2007, MoriYamaguchi:2014} were understood to be related to the Wilson loop operators in the lower 5- and 4-dimensional gauge theories charged in the large rank \textit{anti-symmetric} representations of the $SU(N)$ gauge group. The holographic duals of such Wilson loops in the 4d gauge theories were D5 branes, which were analyzed in detail in \cite{Yamaguchi:2006, GomisPasserini:2006}. The world volume metric of those D5 branes were of $AdS_2 \times S^4$ type which hit the $AdS$ boundary on the location of these Wilson lines. The authors in \cite{Lunin:2007, Chen:2007} discuss an upper bound on the value of fluxes due to 3-form field strength that can be allowed for probe M5 branes for such kind of embedding solutions in $AdS_7 \times S^4$ spacetime geometry. In here, we have also found such a bound in the value of constant $a$ in the 3-form field strength in \eqref{genselfdh}, which will reflect that there is also an upper bound on the flux value on the world volume. \\

\noindent
The first two solutions that we found had the metric in the static gauge with the following identification of the coordinates:
\begin{align}
\label{id1}
 \tau \rightarrow  \, \phi_0, \, \sigma_1 \rightarrow \rho, \, \sigma_2 \rightarrow \alpha, \, \sigma_3 \rightarrow \chi, \, \sigma_4 \rightarrow \xi_1\, \, \sigma_5 \rightarrow \xi_2 \,. 
\end{align}
%
The induced worldvolume metric is of this form
\begin{align}
 ds^2 \Big|_{\text{ind}} \, = \,  4l^2 \left( - \cosh^2 \rho \, d\phi_0^2 \, + \, d\rho^2 \, + \, \sinh^2 \rho \, d\alpha^2 \right) \, + \, l^2 \sin^2 \theta_0 \left( d\chi^2 \, + \, \cos^2 \chi \, d\xi_1^2 \, + \, \sin^2 \chi \, d \xi_2^2  \right)
\end{align}
The $S^4$ coordinate $\theta$ is a constant $\theta_0$ that can take values between $0$ and $\pi$ and also controls the size of $S^3$ wrapped by the worldvolume. Here the $\kappa$ symmetry constraint equation \cite{Howeetal:1997, PSTetal:1997} is of this form
\begin{align}
		\Gamma_{\kappa} \epsilon = \pm \epsilon
\end{align}
where 
\begin{align}
\label{Gkappafluxed}
\Gamma_{\kappa}  =  \frac{1}{\sqrt{- \text{det} \, g}} \left( \gamma_{\tau12345}  \,+\, \frac{40}{6!}  \varepsilon^{mnpqrs} \gamma_{mnp} h_{qrs} \right)
\end{align}
$\gamma_{\tau12345}$ is the product of six worldvolume $\Gamma$ matrices and the indices $m,n,p,q,r,s$ are the coordinate indices. $\varepsilon$ is the Levi-Civita 6-tensor. Here
\begin{align}
		\gamma_{\tau12345}  =  2 l^6 \, \sinh 2\rho \, \sin 2\chi \left( \cosh \rho \, \Gamma_{01289\underline{10}} \, + \, \sinh \rho \, \Gamma_{12489\underline{10}} \right)
\end{align}
matrices $\Gamma$ here are the 11-dimensional matrices with the tangent space indices. \\

\noindent
The non-zero components of the 3-form field strength are
\begin{align}
		h_{\tau\rho\alpha} =  - 4 a l^3 \, \sinh 2\rho  \qquad h_{\chi\xi_1\xi_2} = \frac{a  l^3}2 \, \sin^3 \theta_0 \, \sin 2\chi \,.
\end{align}
which is obtained with the assumption that h obeys
\begin{align}
		\label{sizedependnth}
		h = 2a \left(  \frac{1 + \star_g }2 \right) \Omega_3
\end{align}
with $\Omega_3$ being the volume form on the $S^3$. And therefore, $\Gamma_{\kappa} $ is equal to 
\begin{align}
		= & \frac1{\sqrt{- \text{det} g}} \left( \gamma_{\tau12345}  \,+\, \gamma_{\tau12}\, h_{\chi\xi_1\xi_2}  \,-\, \gamma_{345} \, h_{\tau\rho\alpha}  \right) \cr
		= &  \, \cosh \rho \, \Gamma_{01289\underline{10}} \, + \, \sinh \rho \, \Gamma_{12489\underline{10}} \, + \, a \left( \cosh \rho \, \Gamma_{012} \, + \, \sinh \rho \, \Gamma_{124} \, + \, \Gamma_{89\underline{10}} \right) \,.
\end{align}
In the $\kappa$-symmetry equation, after we commute all the 6-product and 3-product $\Gamma$ matrices(in $\Gamma_{\kappa}$) through the factor $M$ in the killing spinor \eqref{adskss2} we get the following
\begin{align}
\label{solution12gk}
\Gamma_{\kappa} \, M  \, =& \, M \, \bigg[  - \cos \theta_0 \left( e^{ - \Gamma_{78} \chi } e^{- \Gamma_{9\underline{10}} \xi_1 } \Gamma_{89\underline{10}} \left( \cos \beta e^{\Gamma_{14} \phi_1} e^{\Gamma_{25} \phi_2} \Gamma_{012} + \sin \beta e^{\Gamma_{14} \phi_1} e^{\Gamma_{36} \phi_3} \Gamma_{013} \right)   \right)\cr
&\,\,\,  + \sin \theta_0 \left( \cos \beta e^{\Gamma_{14} \phi_1} e^{\Gamma_{25} \phi_2} \Gamma_{012} + \sin \beta e^{\Gamma_{14} \phi_1} e^{\Gamma_{36} \phi_3} \Gamma_{013} \right)  \cr 
&\,\,\, + a \cos \theta_0 \left( \cos \beta e^{\Gamma_{14} \phi_1} e^{\Gamma_{25} \phi_2} \Gamma_{012} + \sin \beta e^{\Gamma_{14} \phi_1} e^{\Gamma_{36} \phi_3} \Gamma_{013} \right) \cr 
& \,\,\, + a \, e^{ - \Gamma_{78} \chi } e^{- \Gamma_{9\underline{10}} \xi_1 } \Gamma_{89\underline{10}} \left( \sin \theta_0 \left( \cos \beta e^{\Gamma_{14} \phi_1} e^{\Gamma_{25} \phi_2} \Gamma_{012} + \sin \beta e^{\Gamma_{14} \phi_1} e^{\Gamma_{36} \phi_3} \Gamma_{013} \right) \,+\, 1 \right) \bigg] 
\end{align}
There are two solutions possible here that satisfy the $\kappa$-constraint
%
%
\begin{enumerate}
	\item We set $\beta$ $=0$, $\phi_1=0$ and $\phi_2=0$ in \eqref{solution12gk} as the required embedding conditions (whereas $\phi_3$ angular direction shrinks to zero size and $\theta$ coordinate has already been set to a fixed $\theta_0$), we find that the worldvolume supersymmetry is preserved if the following projection condition is there
	\begin{align}
	\label{projection1}
		\Gamma_{012} \, \epsilon_0 \, =  \, \epsilon_0 \,,
	\end{align}
	and the value of the constant set to
	\begin{align}
	\label{constanta}
	 \, a = \frac{1 - \sin \theta_0}{\cos \theta_0}\,.
	\end{align}
	\item And on setting $\beta$ $=\frac{\pi}2$, $\phi_1=0$ and $\phi_3=0$ in \eqref{solution12gk} as the embedding conditions (and $\phi_2$ angular direction is of zero size), we find that the worldvolume solution is half-BPS with the following projection condition
	\begin{align}
	\label{projection2}
	\Gamma_{013} \, \epsilon_0 \, =  \, \epsilon_0 \,
	\end{align}
	with $a$ set as \eqref{constanta}.
\end{enumerate}
\textbf{Discussion}: In the equation \eqref{solution12gk} on the RHS, obtained after commuting through the matrix factor $M$, the parenthesis terms in the first two lines are contributions from the six-product $\gamma_{\tau12345}$, and the terms in the $3rd$ and $4th$ lines are from contributions due to the flux terms in the $\kappa$ symmetry equation. Here the value of the 3-form field depends on the value of $\theta_0$ where the $S^4$ coordinate $\theta$ is fixed. By choice from \eqref{sizedependnth}, $h$ depends on the volume of $S^3$, which varies if the fixed value of $\theta$ coordinate $\theta_0$ is changed. And due to the condition \eqref{constanta}, $a$ is also $\theta_0$ dependent. But when $S^3$ vanishes at $\theta$ equal to $0$, and $\pi$ (where the world volume is not $AdS_3 \times S^3$ anymore), these are no longer M5 brane solutions. Whereas at $\theta$ equal to $\frac{\pi}2$, although the size of the $S^3$ wrapped by the world volume is maximal, $a$ becomes zero here. So in the limit when value of $\theta \rightarrow \frac{\pi}2$, even with flux field $h$ $\rightarrow$ $0$, these are still stable supersymmetric $AdS_3 \times S^3$ probe M5 solutions.
\vspace{.5cm}

There are four more half-BPS solutions that contain such 3-form flux field. Identifications for the M5 worldvolume coordinates of these are different from \eqref{id1}.  $\Gamma_{\kappa}$ can also be computed for all these four solutions separately and an equation similar to \eqref{solution12gk} can be found. The value of the constant $a$ for all these are the same as in \eqref{constanta}.
Below we merely list these solutions with their embedding conditions and along with the respective projection conditions. We find that the three of them can be nicely written in terms of the complex $AdS_7$ coordinates defined in \eqref{complexAdS7}

	\subsubsection*{$\,3:\,\, \Phi_2 = 0$ and $\Phi_3=0$}
	In terms of the real coordinates the embedding conditions are: $\alpha=0$ and $\theta = \theta_0$. When $\alpha=0$ the angular directions parametrized by the coordinates $\beta$, $\phi_2$ and $\phi_3$ shrink to zero sizes. The identification of the worldvolume coordinate in the static gauge is
	$$\left( \tau, \sigma_1, \sigma_2, \sigma_3, \sigma_4, \sigma_5 \right) \,\, \rightarrow \,\, \left( \phi_0, \rho, \phi_1, \chi, \xi_1, \xi_2 \right)$$ 
	This solution is supersymmetric with $\kappa$-symmetry constraint true when
	\begin{align}
	\label{projection3}
		\Gamma_{014} \, \epsilon_0 \, = \,  \epsilon_0 \,.
	\end{align}
	\subsubsection*{$\,4:\,\, \Phi_1 = 0$ and $\Phi_3=0$}
	In terms of the real coordinates the embedding conditions are: $\alpha=\frac{\pi}2$, $\beta = 0$ and, $\theta = \theta_0$. When $\alpha=\frac{\pi}2$ and $\beta = 0$ the angular directions parametrized by coordinates $\phi_1$ and $\phi_3$ shrink to zero size. The identification of the worldvolume coordinates is
	$$\left( \tau, \sigma_1, \sigma_2, \sigma_3, \sigma_4, \sigma_5 \right) \,\, \rightarrow \,\, \left( \phi_0, \rho, \phi_2, \chi, \xi_1, \xi_2 \right)$$ 
	The solution is supersymmetric with $\kappa$-symmetry constraint true when
	\begin{align}
	\label{projection5}
	\Gamma_{025} \, \epsilon_0 \, = \,  \epsilon_0 \,.
	\end{align}
	\subsubsection*{$\,5:\,\, \Phi_1 = 0$ and $\Phi_2=0$}
	In terms of the real coordinates the embedding conditions are: $\alpha=\frac{\pi}2$, $\beta = \frac{\pi}2$ and, $\theta = \theta_0$.  When $\alpha=\frac{\pi}2$ and $\beta = \frac{\pi}2$ the angular directions parametrized by coordinates $\phi_1$ and $\phi_2$ shrink to zero size. The identification of the worldvolume coordinates is
	$$\left( \tau, \sigma_1, \sigma_2, \sigma_3, \sigma_4, \sigma_5 \right) \,\, \rightarrow \,\, \left( \phi_0, \rho, \phi_3, \chi, \xi_1, \xi_2 \right)$$ 
	The solution is supersymmetric when
	\begin{align}
	\label{projection6}
	\Gamma_{036} \, \epsilon_0 \, = \,   \epsilon_0 \,.
	\end{align}
	\textbf{ $6:\,\,$ For the last solution}, the real embedding conditions are: $\alpha=\frac{\pi}2$, $\theta = \theta_0$, $\phi_2 = 0$ and $\phi_3 = 0$. And the $\phi_1$ direction shrinks to zero size. With the following worldvolume identification
	$$\left( \tau, \sigma_1, \sigma_2, \sigma_3, \sigma_4, \sigma_5\right) \,\, \rightarrow \,\, \left( \phi_0, \rho, \beta,  \chi, \xi_1, \xi_2 \right)$$
	This solution is supersymmetric when
	\begin{align}
	\label{projection4}
	\Gamma_{023} \, \epsilon_0 \, = \,  \epsilon_0 \,.
	\end{align}
	%
%
%
\subsection{General \textit{giant-like} solutions in kind I}\label{common susy}

It is also essential to identify the sets among these solutions that have common Supersymmetries
\begin{itemize}
	\item $\frac14^{\text{th}}$ common supersymmetries: 
	\begin{enumerate}
			\item Solution with projection \eqref{projection1} and the solution with projection \eqref{projection6}.
			\item Solution with projection \eqref{projection2} and the solution with projection \eqref{projection5}.
			\item Solution with projection \eqref{projection3} and the solution with projection \eqref{projection4}.
	\end{enumerate}
	Considering the above sets of projections this will give us the individual $\frac14^{\text{th}}$-BPS embedding solutions separately.
	\item $\frac18^{\text{th}}$ common supersymmetries: There is only one combination that is possible with projections of \eqref{projection3}, \eqref{projection5} and \eqref{projection6}  which we also rewrite as follows
 \begin{align}
\label{oneeightBPS}
\Gamma_{14} \, \epsilon_0 \, = \,  \Gamma_{25} \, \epsilon_0 \, = \,	\Gamma_{36} \, \epsilon_0 \, = \, -  \Gamma_0 \, \epsilon_0 \,
\end{align}
along with the $\frac18$-BPS projections in \eqref{oneeightBPS} if we also consider 11-Gamma product $\Gamma_{01234\ldots\underline{10}}$ $=$ $1$ then the following is also true
\begin{align}
\label{oneeightBPSextra}
		\Gamma_{7\underline{10}} \, \epsilon_0 \, = \, - \Gamma_{89} \, \epsilon_0 \,.
\end{align}
\end{itemize}
%
%
%
%
%
In the remainder of the section, we aim to find a general expression solution that will also include the three half-BPS solutions that preserve the common supersymmetry due to \eqref{oneeightBPS}. To find such an expression we further break the supersymmetry by introducing an extra projector
\begin{align}
\label{onesixteenthBPS}
\Gamma_{14} \, \epsilon_0 \, = \,  \Gamma_{25} \, \epsilon_0 \, = \,	\Gamma_{36} \, \epsilon_0 \, = \, -  \Gamma_0 \, \epsilon_0 \, = \, i \, \epsilon_0 \,.
\end{align}
The $\frac 1{16}$-BPS projections above will also preserve the single supercharge that was determined by the $\frac 1{32}$ projections we wrote in \cite{VG:21} to calculate the most general $\frac 1{32}$-BPS solution. $\theta$ equal to $\frac{\pi}2$ was another embedding condition for that solution. We rewrite the most general solution of \cite{VG:21} for the convenience below
\begin{align}
\label{generalsolutions1by32}
F^{(I)} ( \Phi_0 \,, \Phi_1\,, \Phi_2 \,, \Phi_3 \,, Z_1, Z_2) = 0 \qquad \qquad \text{with}\,\, I=1,2\,,
\end{align}
\begin{align}
\label{scalingcondition4gen}
\text{with the scaling condition on each}\,\, F^{(I)}: \hspace{1.8cm} \sum_{i=0}^3 F^{(I)}_{\phi_i} - 2 \sum_{i=1,2} F^{(I)}_{\xi_i} \,= \, 0 \,.
\end{align}
%
%
 %
Since the $\frac 1{16}$-BPS solution, for the \textit{giant-like} M5 brane that we are after, will also preserve the single supersymmetry of \eqref{generalsolutions1by32}. This will mean that it will be contained(as a subset conditions) within the \eqref{generalsolutions1by32} expression. But because it preserves at least one more supersymmetry than \eqref{generalsolutions1by32} the solution expression will be of a more constrained form. The form of our $\frac1{16}$-BPS solution here will be different from the ones in equation (4.1) of \cite{VG:21} where one of the $Z$ coordinates of $S^3$ $\subset$ $S^4$ was put to zero. We rewrite this particular
 equation as well
 \begin{align}
 \label{1by16BPSsolutionform}
 Z_2 = 0 \quad \text{and} \quad F\left( \zeta \Phi_0 , \zeta \Phi_1, \zeta \Phi_2, \zeta \Phi_3 \right) = 0 \,
 \end{align}
  where $\zeta = \sqrt{Z_1} = e^{i \frac{\xi_1}2 }$. This equation is of further constrained form from \eqref{generalsolutions1by32} that goes on to describe the holographic duals of codimension-2 defects in the boundary 6d SCFT. This above equation  of \cite{VG:21}, now labeled \eqref{1by16BPSsolutionform}, describes another $\frac1{16}$-BPS sector of M5 brane solutions different from our current $\frac1{16}$-BPS sector here due to \eqref{onesixteenthBPS}.
M5 worldvolume of \eqref{1by16BPSsolutionform} wraps an arbitrary 1d curve on the $S^3$ $\subset$ $S^4$ and the remaining $1+4$ directions of the 6d worldvolume lie completely in the $AdS_7$ part.
While in the present case, the 3-dimensional space due to sphere $S^3$ $\subset$ $S^4$ wrapped by a \textit{giant-like} M5 solutions is always going to be the same, and hence the condition $Z_2 =0$ will not enter here. 
  \\

 It is also easy to illustrate how the individual half-BPS solutions corresponding to \eqref{projection3}, \eqref{projection5} and \eqref{projection6} will fit into the general form of \eqref{generalsolutions1by32}. For example, the half-BPS solution due to \eqref{projection3} can also be expressed as 
 \begin{align}
		\Phi_2 = 0 \quad \qquad \Phi_3 = 0 \,.
 \end{align}
Hence we can take the holomorphic functions in \eqref{generalsolutions1by32} to be
 \begin{align}
 	&	F^{(1)} ( \Phi_0 \,, \Phi_1\,, \Phi_2 \,, \Phi_3 \,, Z_1, Z_2)  \,=\,  \Phi_2 \,=\, 0   \cr
 		&	F^{(2)} ( \Phi_0 \,, \Phi_1\,, \Phi_2 \,, \Phi_3 \,, Z_1, Z_2)  \,=\,  \Phi_3\,=\, 0 \,.
 \end{align}
 Now, since the worldvolume wrap all of $S^3$ $\subset$ $S^4$, coordinates $Z_1$, $Z_2$ will not appear as arguments. Therefore, it is natural to conclude that the answer for general $\frac1{16}$-BPS M5 solution of this \textit{``giant-like"} kind can be expressed in terms of two arbitrary holomorphic conditions:
 \begin{align}
 \label{newgensolution}
F^{(I)} ( \Phi_0 \,, \Phi_1\,, \Phi_2 \,, \Phi_3 )  \,=\,   0   \qquad \text{with} \,\, I =1,2 \,,
 \end{align}
 along with the scaling condition
\begin{align}\label{scalingcondition}
   \sum_{i=0}^3 \partial_{\xi_i} F^{(I)} ( \Phi_0 \,, \Phi_1\,, \Phi_2 \,, \Phi_3) = 0
\end{align}
with the coordinate $\theta$ fixed to any arbitrary value $\theta_0$ between $0$ and $\pi$. However, the most general solution \eqref{generalsolutions1by32} from \cite{VG:21} did not have the 3-form field strength of the M5 world volume turned on. \par 
%
%

For completion, we give explicit proof for the general expression \eqref{newgensolution} in appendix \ref{AppA}. It does not involve any new detail(method-wise) that we have not discussed so far here and in \cite{VG:21}.

 \subsection{Boundary limit description} 
 
 We can look at this general solution in the limit where the $AdS$ radius is large. The complex coordinates $\Phi_i \in \mathbb{C}^{1,3}$ chosen for the $AdS_7$ embedding take the form
 \begin{align}
 		\Phi_0 \, = \, r \nu_0 \qquad  \Phi_1 \, = \, r \nu_1 \qquad \Phi_2 \, = \, r \nu_2 \qquad \Phi_3 \, = \, r \nu_3 
 \end{align}
 where $r = 2 l \, \sinh \rho$ and $\nu_0 = e^{i\phi_0}$, $\nu_1 = \cos \alpha e^{i \phi_1}$, $\nu_2 = \sin \alpha \cos \beta e^{i \phi_2}$, $\nu_3 = \sin \alpha \sin \beta e^{i \phi_3}$. In this large r approximation(where the boundary theory lies), the induced metric is in the conformal class of $\mathbb{R} \times S^5$ 
 \begin{align}
 - |d\Phi_0|^2 + |d\Phi_1|^2 + |d\Phi_2|^2 &+ |d\Phi_3|^2 = \cr 
 &r^2 \left( - d\phi_0^2 + d\alpha^2 + \cos^2 \alpha \, d\phi_1 + \sin^2 \alpha \left( d\beta^2 + \cos^2 \beta \, d\phi_2^2 + \sin^2 \beta \, d\phi_3^2 \right) \right)  \,. \cr
 \end{align}
 Now we need to find the locus of zeros of the functions in \eqref{newgensolution} as we near the boundary. It is important to note that the 
 holomorphic functions $F^{(I)}$ given in \eqref{newgensolution} with a differential scaling condition given in \eqref{scalingcondition} can be repackaged in the following way
 \begin{align}
   F^{(I)} ( \Phi_0 \,, \Phi_1\,, \Phi_2 \,, \Phi_3 ) \,\, \approx \,\, H^{(I)} \left( \frac{\Phi_1}{\Phi_0}, \, \frac{\Phi_2}{\Phi_0}, \, \frac{\Phi_3}{\Phi_0} \right) \,=\, 0
 \end{align}
 where $H^{(I)}$ are two arbitrary holomorphic functions. Near the boundary the functions become $H^{(I)}\left(  \frac{ \nu_1}{\nu_0}, \, \frac{\nu_2}{\nu_0}, \,  \frac{\nu_3}{\nu_0}  \right)$. So the worldvolume of the M5 brane intersects the boundary(at radial coordinate $r=r_c$) at the common intersection of zeros of the functions
 \begin{align}\label{codimension4location}
 H^{(I)}\left(  \frac{ \nu_1}{\nu_0}, \, \frac{\nu_2}{\nu_0}, \,  \frac{\nu_3}{\nu_0}  \right) = 0 \,.
 \end{align}
 %
 %
 %
 \noindent
 Next, we show how to connect this general solution with a class of solutions that appeared for Wilson surface operators in \cite{Drukkeretal:1120}. It can be seen that if we shrink the size of the $S^3$ wrapped by the worldvolume(by moving to the $\theta$ position $\theta=0$) the resulting M2 branes are valid solutions preserving the same supersymmetry before the general M5 brane becomes degenerate, in this class of solutions. The M2 brane embedding will be described by the same conditions in \eqref{newgensolution} and will not wrap any directions contained in the $S^4$ part of the 11d background geometry. \\

\noindent
 First we do the Wick rotation $\tau_E \, = \, - i \, \tau$, the functional conditions of \eqref{codimension4location} take the form
 \begin{align}
 			\label{functionalcond}
 		 H^{(I)}\left( e^{ \tau_E}  \nu_1 , \, e^{ \tau_E}  \nu_2 , \, e^{ \tau_E}   \nu_3 \right) \,=\, 0 \,.
 \end{align}
 Then we do the conformal coordinate transformation: $$(z_1, \, z_2,\, z_3) =  e^{\tau_E} \left( \cos \alpha \, e^{i \phi_1},\, \sin \alpha \cos \beta \,  e^{i \phi_2}, \, \sin \alpha \sin \beta \, e^{i \phi_3} \right)$$
 \vspace{.5cm}
 The metric transforms as follows
 \begin{align*}
  \, & d\tau_E^2 + d\alpha^2 + \cos^2 \alpha \,  d\phi_1^2  \vertarrowbox{+ \sin^2 \alpha}{}\left( d\beta^2 + \cos^2 \beta \, d\phi_2^2 + \sin^2 \beta \, d\phi_3^2  \right)  \cr
  & e^{2\tau_E} \, \left(  d\tau_E^2 + d\alpha^2 + \cos^2 \alpha \,  d\phi_1^2  + \sin^2 \alpha \left( d\beta^2 + \cos^2 \beta \, d\phi_2^2 + \sin^2 \beta \, d\phi_3^2  \right) \right) \, \,= \,\,\,\,  |dz_1|^2 \, + \,  |dz_2|^2 \, + \,  |dz_3|^2
 \end{align*}
  It can now be seen that the functional conditions in \eqref{functionalcond} describe an arbitrary holomorphic curve in the $\mathbb{C}^3$ space
 \begin{align}
 \label{SigmainC3}
 H^{(I)}\left(z_1 ,\, z_2,\, z_3\right) \,=\, 0 \, \qquad \qquad \text{for} \,\, I=1,2\,.
 \end{align}
 Therefore, we have shown that after we make our general $\frac 1{16}$ BPS M5 brane solution degenerate to M2 branes of the same supersymmetry(after shrinking the $S^3$ size to 0), in the large AdS radius boundary limit, our derived functional conditions in \eqref{functionalcond} describe an arbitrary holomorphic curve in the $\mathbb{C}^3$ space, and which give the `type-C' Wilson surfaces obtained in the reference \cite{Drukkeretal:1120}. 
 

The direct one-to-one map to the solutions of \cite{Drukkeretal:1120} also makes sense with the understanding that the `scalar couplings' $n^{I}$, which are defined in \cite{Drukkeretal:1120}, were constant vectors along the whole of Wilson surfaces only for the `type-C' class surfaces. The $n^{I}$ vectors were unit vectors that took values in $\mathbb{R}^5$, the same space in which the 5 scalar fields of the boundary theory take their values. For our general solutions from \eqref{newgensolution}, the associated $n^{I}$ vectors for the boundary scalars are going to be constants by construction, since the holographic dual M5 solutions always wrap all of the $S^3$ directions inside $S^4$ (which are the orthogonal directions to the $AdS_7$ boundary) and are located at fixed $\theta$ coordinate values. And in the degenerate limit to get M2 branes $\theta=0$, where $S^3$ size is zero.\\

 \noindent
 \textbf{Some comments}: It is also worthwhile to note that because $S^4$ coordinates: $Z_1$ and $Z_2$ are not the arguments of these holomorphic conditions(as they are all identified with the worldvolume coordinates that describe the $S^3$ part wrapped by the general M5 solution), they do not appear as arguments in the general embedding conditions we have presented in \eqref{newgensolution}.
 Due to this, we cannot do the analysis that we did for the profile of the scalar fields singularities associated with the codimension-2 defects in the later part of section 4 in \cite{VG:21}, where the coordinate $Z_1$ was still in the argument (refer \eqref{1by16BPSsolutionform}). And we had taken the boundary limit for the $AdS$ radial direction and then found the singularity profile of the complex scalar in the boundary $(2,0)$ tensor multiplet theory.

 Therefore, for these \textit{giant-like} M5 branes, it appears that unlike the previous solutions in \cite{VG:21}, taking the large radius limit will not reflect any singular behaviour for the boundary gauge theory scalar fields(when the codimension-4 defects are present). This is very much consistent with what is known about surface operators and also discussed in \cite{FGT:2015, Gukov:2014} where, upon dimensional reduction, the codimension-4 defects were associated with the coupled 2d/4d description of surface operators in the 4-dimensional gauge theory.

 %
\section{Solutions of the kind II:  $\tilde{S}^3$ $\subset$ $AdS_7$}

The \textit{dual-giant-like} solutions that have been known are also understood to be related to those Wilson operators which are charged in the large rank \textit{symmetric} representations of the $SU(N)$ gauge group of the 4- and 5-dimensional gauge theories. The holographic dual probe D3 branes in the $AdS_5 \times S^5$ spacetime were analyzed in \cite{Drukker:2005kx, GomisPasserini:2006} where world volume metric is of $AdS_2 \times S^2$ type and hit the $AdS$ boundary on the location of such Wilson lines.
For the \textit{dual-giant-like kind} that we present, there are three half-BPS solutions preserving supersymmetries in the different sectors. Each has a respective projection condition on the constant spinor $\epsilon_0$ required to satisfy the $\kappa$-symmetry constraint. They all have the following identification for the worldvolume coordinates
\begin{align}
\label{id7}
\tau \rightarrow \phi_0, \, \sigma_1 \rightarrow \rho, \, \sigma_2 \rightarrow \phi_1, \, \sigma_3 \rightarrow \beta, \, \sigma_4 \rightarrow \phi_2\,, \, \sigma_5 \rightarrow \phi_3 \,. 
\end{align}
The three M5 brane solutions that we discuss here directly couple to the dual background 6-form potential field
\begin{align}
C^{(6)} = - \left( 2 l\right)^6  \sinh^6 \rho  \cos \alpha  \sin^3 \alpha  \cos \beta  \sin \beta \, d\alpha \wedge d\beta \wedge d\phi_0 \wedge \ldots \wedge d\phi_3 ~.
\end{align}
The embedding condition: $\sinh \rho \sin \alpha = 1$ is also the same for all three. The induced worldvolume metric of $AdS_3 \times S^3$ topology has the following expression
\begin{align}
 ds^2 \, =\, 2 l^2\left( -2 \, \cosh^2 \rho \, d\phi_0^2 \, + \, \frac{4\, d\rho^2 }{1 - \text{csch}^2 \rho} \, + \,  \left( \cosh 2\rho - 3 \right) d\phi_1^2 \right) \, + \, 4l^2 \left(  d\beta^2 \, + \, \cos^2\beta d\phi_2^2 \, + \, \sin^2\beta d\phi_3^2 \, \right)
\end{align}
What is different are the following embedding conditions
\begin{itemize}
	\item \textbf{Solution 1}: the point chosen is $\theta=\frac{\pi}2$, $\chi=0$ and $\xi_1 = 0$ where the circle parametrized by $\xi_2$ shrinks.
	\item \textbf{Solution 2}: the point chosen is $\theta=\frac{\pi}2$, $\chi=\frac{\pi}2$ and $\xi_2 = 0$ where the circle parametrized by $\xi_1$ shrinks.
	\item \textbf{Solution 3}: the point $\theta=0$ is chosen where the size of $S^3$ $\subset$ $S^4$ shrinks to zero.
\end{itemize}
We demonstrate our calculation a bit for \textbf{Solution 1} here and later state the final answer for all of them. For convenience we write the $\Gamma_{\kappa}$ matrix in its general form again:
\begin{align}
\label{Gkappafluxedagain}
\Gamma_{\kappa}  =  \frac{1}{\sqrt{- \text{det} \, g}} \left( \gamma_{\tau12345}  \,+\, \frac{40}{6!}  \varepsilon^{mnpqrs} \gamma_{mnp} h_{qrs} \right)
\end{align}
For the chosen solution the matrix $\frac{1}{\sqrt{- \text{det} \,g}}\gamma_{\tau12345}$ is equal to
\begin{align}
\label{kind2gammak}
	-\frac1{\sqrt{2}} \left( \coth \rho \Gamma_{023456} - \sqrt{1 - \sinh^{-2} \rho} \, \Gamma_{013456} \right)
\end{align}
and the flux-dependent part is equal to 
\begin{align}
\label{kind2Gflux}
   & \frac{1}{\sqrt{- \text{det} \, g}} \left( \gamma_{\tau12} \, h_{\beta\phi_2 \phi_3}  \, - \, \gamma_{345} \, h_{\tau\rho\phi_1}  \right)\cr 
   = &  - \frac{a}{\sqrt{2}}  \bigg[ \sqrt{1 - \sinh^{-2} \rho}  \, \Gamma_{145} + \left(1 - \sinh^{-2} \rho \right) \, \Gamma_{014} \cr 
   &- \cot \rho \left( \Gamma_{245} -  \cot \rho \, \Gamma_{025}  +  \sqrt{1 - \sinh^{-2} \rho}  \left( \Gamma_{015} + \Gamma_{024} \right)\right) \bigg] \cr 
   & + \frac{a}{\sqrt{2}} \left[ \coth \rho \, \Gamma_{346} - \left( \Gamma_{036} - \sqrt{1 - \sinh^{-2} \rho} \Gamma_{356} \right) \right]
\end{align}
Here the non-zero components of the self-dual field strength are
\begin{align}
	h_{\tau\rho\phi_1} = \, - 4 a \,  l^3 \sqrt{\cosh 4\rho - 1} \qquad \quad h_{\beta\phi_2\phi_3} = 4 a \, l^3 \sin 2 \beta
\end{align}
After commuting all the six-product $\Gamma$ matrices in \eqref{kind2gammak} and all the 3-product $\Gamma$ matrices in \eqref{kind2Gflux} through the matrix factor $M$ in the killing spinor $\epsilon$ \eqref{adskss2} we find that the $\kappa$-symmetry constraint is satisfied. \\

\noindent
The projection required on the constant spinor $\epsilon_0$ is the following
\begin{align}
 \Gamma_{01489\underline{10}} \, \epsilon_0  \, = \,  \epsilon_0 \,.
\end{align}
Along with this, we find that the value of the constant factor $a$ in the $h$ needs to be 
\begin{align}
		a  \, = \, 1  - \sqrt{2}
\end{align}
Similarly, the answers for the remaining two solutions are \\

\noindent
\textbf{- Solution 2 :} with the real embedding conditions $\theta=\frac{\pi}2$, $\chi=\frac{\pi}2$ and $\xi_2 = 0$
	\begin{align}
	\Gamma_{01479\underline{10}} \, \epsilon_0  \, = \, - \epsilon_0  
	\end{align}
\textbf{- Solution 3 :} with $\theta=0$ condition
 \begin{align}
		\Gamma_{014} \, \epsilon_0 =   \epsilon_0 
 \end{align}
%
Here the contributions from both $\gamma_{\tau12345}$ and the flux-dependent part in the $\Gamma_{\kappa}$ matrix \eqref{Gkappafluxedagain} are needed in the $\kappa$-symmetry equation for the solutions to be supersymmetric. Therefore, we find that for these half-BPS solutions the value of 3-form flux field $h$ cannot vanish in order to have a supersymmetric worldvolume of $AdS_3 \times S^3$. The same conclusion was also found for the \textit{"dual-giant-like"} solutions in \cite{Chen:2007}.

This also means that it will not be possible for these solutions to be combined with those that we saw in the previous section. Hence they will not belong to the same BPS sector and they will not belong to the embedding conditions in \eqref{generalsolutions1by32} as done by those \textit{giant-like} kinds and also by the solutions in \cite{VG:21}.

\section{Discussion of the results and conclusion}\label{conclusionsection}
In this article, we presented some probe M5 branes in the $AdS_7 \times S^4$ spacetime with half-BPS solutions having $AdS_3 \times S^3$ world volume. These solutions have the world volume 3-form flux field turned on. Two kinds of solutions appeared in our discussion i) \textit{giant-like} solutions and ii) \textit{dual-giant-like} solutions. Both the kinds of M5 solutions will hit the boundary region of $AdS_7$ at the locations of $1 + 1$ dimensional Wilson surface operators in the dual gauge theory. From supersymmetry analysis, we found these solutions to be consistent with the ones discussed earlier in the literature \cite{Lunin:2007, Chen:2007, MoriYamaguchi:2014}.
For the \textit{giant-like} solutions, we see that the worldvolume 3-form flux field value becomes maximum when the M5 brane is located at a fixed point of $\theta_0$, which is somewhere between $0$ and $\frac{\pi}2$. An increment of $\theta_0$ from this 'maximum point' decreases the value of the flux field before it vanish at $\theta_0 = \frac{\pi}2$. 
\begin{center}
	\begin{figure}[h]
		\begin{center}\includegraphics[width=40pc]{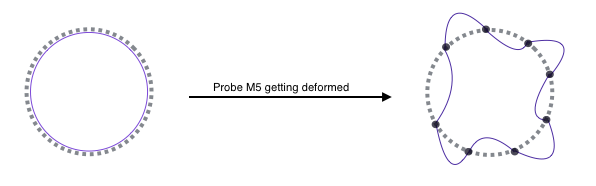} \end{center}
		\caption{{ \label{deformedM5} This figure depicts the probe M5 brane dual to codimension-2 defects in the background spacetime geometry of $AdS_7 \times S^4$ getting deformed to become the M5 which is dual to codimension-4 defects.
				\footnotesize   
		}}
	\end{figure}
\end{center}
After this, we found that 3 of the half-BPS \textit{giant-like} solutions can be combined such that a general solution can be constructed that preserves the common four supersymmetries of these. On further breaking the supersymmetry by an extra $\frac12$ factor, we found that the common supersymmetries due to the 4 independent projections in equation \eqref{onesixteenthBPS} were among the $\frac1{32}$-BPS projections we found in \cite{VG:21}. In equation \eqref{newgensolution}, we also determine the expression for a more general $\frac1{16}$-BPS solution that can have the 3-form flux field turned on. By AdS/CFT holographic principle, this general solution will describe the probe M5 branes that are dual to those $\frac1{16}$-BPS codimension-4 surface defects in the 6d boundary gauge theory which would be charged in the \textit{higher rank representation} of the gauge symmetry. Towards the end of section 3, we further show how the general solution in \eqref{newgensolution} is mapped to the `type-C' Wilson surfaces of \cite{Drukkeretal:1120} (refer to \cite{Wangetal:2018} for surfaces of other general shapes and classes).
In \cite{VG:21}, we also found a solution for the M5 duals of $\frac1{16}$-BPS codimension-2 defects in the gauge theory, which we write in \eqref{1by16BPSsolutionform}. In this concluding section, we would like to end with some comments on the relation between these two different $\frac1{16}$-BPS solutions. In the $AdS_7 \times S^4$ background spacetime, the brane construction of these can be understood from the table below
 \begin{table}[H]
	\begin{center}
		\begin{tabular}{|c|c|c|c|c|c|c|c|c|c|c|c|}
			\hline
			 &  $0$ & $1$& $2$& $3$& $4$& $5$& $6$ & $7$ &$8$ & $9$ & $10$ \cr
			\hline
			M$5_N$ &  $\times$& $\times$& $\times$& $\times$& $\times$& $\times$& $-$ & $-$ &$-$ & $-$ & $-$ \cr
			\hline
			M$5_{\text{codim-2}}$ &  $\times$& $\times$& $\times$& $\times$& $-$ & $-$ &  $\times$ & $-$ & $-$ & $-$& $\times$ \cr
			\hline
			M$5_{\text{codim-4}}$ &  $\times$& $\times$&  $-$ & $-$ & $-$ &  $-$ & $\times$ &  $-$& $\times$& $\times$& $\times$\cr
			\hline
		\end{tabular}
	\end{center}
	\caption{The construction of the probe M5 duals of codimension-2 and codimension-4 defects in the background $AdS_7 \times S^4$ spacetime geometry can be understood as above. } 
	\label{M5construction}
\end{table}
In \cite{FGT:2015}, authors discussed this relationship between these two kinds of probe M5 branes in terms of deformation, such that M$5_{\text{codim-2}}$ brane that intersects with the boundary region of $AdS_7$ in four directions, can be deformed to an M$5_{\text{codim-4}}$ brane which has just two common directions with the $AdS$ boundary (also see fig. \ref{deformedM5}). For us this kind of deformation doesn't change the number of supersymmetry preserved, but the deformed version, the M$5_{\text{codim-4}}$, now belong to a different $\frac1{16}$-BPS sector of the solutions. However, for the construction of Frenkel, Gukov and Teschner in \cite{FGT:2015}, the BPS sector also remained the same after such worldvolume deformation. \par

\noindent
For our case, the equation below captures this deformation process in a precise way 
\begin{align*}
& M5_{codim-2} \,\,\, :\, \,\,\vertarrowbox{F \left( Z \Phi_0 , Z \Phi_1 , Z \Phi_2 , Z \Phi_3 \right)}{on deformation}  = 0   \quad -- \quad  \text{World volume flux zero;} \,\,\, \theta \subset S^4  =  \frac{\pi}2 \,; \cr
&M5_{codim-4} \,\, \,:\, \,\, F^{(I)} \left( \Phi_0 ,  \Phi_1 ,  \Phi_2 ,  \Phi_3 \right) = 0 \quad -- \quad  \text{World volume flux vanish as $\theta \rightarrow \frac{\pi}2$}  
\end{align*}
The authors in\cite{FGT:2015} also discussed Hanany-Witten-like phenomenon \cite{HananyWitten:96} when such deformation of M5 brane was considered. They show the emergence of M2 branes when the deformed M$5_{\text{codim-4}}$ brane is moved in the direction that is orthogonal to all the three kinds of M5s shown in table \ref{M5construction}.
But here in the current scenario, such M2 branes, if present, should be stretched along the radial $AdS_7$ direction and also intersect the boundary at the same $1+1$ dimensional location where M$5_{\text{codim-4}}$ brane is hitting. But the presence of these M2 branes is obscured at the moment.
Perhaps, a close analysis of the movement of M$5_{\text{codim-4}}$ brane along the $\theta  \subset S^4$ coordinate will help us in recovering them. We will look forward to addressing this shortly. Another direction that one can pursue is to also turn on the flux field $h$ on the worldvolume of M$5_{\text{codim-2}}$ branes from reference \cite{VG:21}. Our analysis for M$5_{\text{codim-4}}$ branes in this article suggests that changing the position of M$5_{\text{codim-2}}$ from $\theta = \frac{\pi}2$ to an arbitrary $\theta_0$ may allow for such a possibility. Doing this may also shine more light on the intricacies in the relation between the two descriptions of surface defects(which are of codimension-2 and codimension-4) in the six-dimensional boundary gauge theories.
	\vspace{.9cm}
\noindent
\textbf{\normalsize Acknowledgements}: 
 \small{We are grateful to Sujay Ashok for the support and many discussions throughout this work. We thank Nemani V. Suryanarayana for valuable comments and discussions on the work that appeared here. We also like to thank K. Narayan for many related discussions. We also thank Nadav Drukker and Maxime Tr\'epanier for their comments on an earlier version of the article. This work is partially supported by a grant to CMI from the Infosys Foundation.}

    \vspace{.5cm}

 \appendix

\section{Choice of frame vielbeins}

The global $AdS_7$ coordinates in the equation \eqref{gddbulk} can be written in terms of the following complex coordinates in $\mathbb{C}^{1,3}$
		\begin{align}
		\label{complexAdS7}
		\Phi_0 = l \cosh \rho  \, e^{i \phi_0} \quad \Phi_1 = l \sinh \rho \cos \alpha \, e^{i \phi_1} \quad  \Phi_2 = l \sinh \rho \sin \alpha \cos \beta \, e^{i \phi_2} \quad  \Phi_3 = l \sinh \rho \sin \alpha \sin \beta \, e^{i \phi_3}
		\end{align}
		%
		%
		%
		For the $S^3$ $\subset$ $S^4$ we define the complex cooordinates describing it embedded in $\mathbb{C}^2$ space
		\begin{align}
		Z_1 = \cos \chi \, e^{i \xi_1}  \quad \qquad \quad  Z_2 = \sin \chi \, e^{i \xi_2} \,.
		\end{align}
		%
		We choose the same frame vielbein that we had taken in \cite{VG:21} where the $AdS_7$ part can be written as a $U(1)$ Hopf fibration over a K\"ahler manifold $\widetilde{\mathbb{CP}}^3$ (also refer to \cite{AGS:2020}). Here $\widetilde{\mathbb{CP}}^3$ is the hyperbolic version of the complex projective space and it is defined as the set of rays in the complex space $\mathbb{C}^{1,3}$( in place of the $\mathbb{C}^4$ space). Similarly for the $S^3 \subset S^4$ part, the frame vielbein are chosen so that $U(1)$ Hopf fibration over a K\"ahler manifold $\mathbb{CP}^1$ becomes manifest.
		The frame vielbein that we use are the following
		\begin{align}\label{adsframe}
		e^0 &= 2 l \left( \cosh^2 \rho \, d\phi_0 - \sinh^2 \rho \left( \cos^2 \alpha \, d\phi_1 + \sin^2 \alpha \cos^2 \beta \, d \phi_2 + \sin^2 \alpha \sin^2 \beta \, d \phi_3 \right)\right) \cr
		e^1 &= 2 l \, d \rho , \quad e^2 = 2 l  \sinh \rho \, d\alpha , \quad e^3 = 2 l \sinh \rho \sin \alpha \, d\beta \cr 
		e^4 &= 2 l \cosh \rho \sinh \rho \left( \cos^2 \alpha \, d\phi_{01} + \sin^2 \alpha \cos^2 \beta \, d\phi_{02} + \sin^2 \alpha \sin^2 \beta \, d\phi_{03} \right)   \cr
		e^5 &= 2 l \sinh \rho \cos \alpha \sin \alpha \left( \cos^2 \beta \, d\phi_{02} + \sin^2 \beta \, d\phi_{03} - d\phi_{01} \right) \cr
		e^6 &= 2 l \sinh \rho \sin \alpha \cos \beta \sin \beta \left( d\phi_{03} - d\phi_{02} \right)
		\end{align}
		%
		where $r = 2 l \sinh \rho$, $\phi_0 = \frac t{2l}$ (the notation $d\phi_{0i}$ here just means $d\phi_{0}  -  d\phi_i$), and
		\begin{align}\label{S4frame}
		e^7 &= l \, d\theta , \quad e^8 = l\,\sin \theta  d\chi  , \quad e^9 = l \,\sin \theta \cos \chi \sin \chi \left(  d\xi_1 \, -  \, d\xi_2 \right) \,, \cr 
		e^{\underline{10}} &= l\,\sin \theta \left( \cos^2 \chi \, d\xi_1 + \sin^2 \chi \, d\xi_2 \right) \,.
		\end{align}
		\noindent
		%
		%
		%
		%
		%
		%
		The solution for the Killing spinor in the above coordinate is given by :
		\begin{align}
		\label{adskss2}
		\epsilon &=  e^{\frac 12 ( \Gamma_{04} + \Gamma_1 \gamma ) \rho} e^{\frac 12 \left( \Gamma_{12} + \Gamma_{45} \right) \alpha} e^{\frac 12 \left( \Gamma_{23} + \Gamma_{56} \right) \beta}  e^{\frac 12 \Gamma_0 \gamma  \phi_0} e^{ - \frac 12 \Gamma_{14} \phi_1 } e^{ - \frac 12 \Gamma_{25} \phi_2} e^{ - \frac 12 \Gamma_{36} \phi_3} \cr
		& \hspace{1cm} \times e^{\frac 12 \gamma \Gamma_7 \theta } e^{ \frac 12 \left( \Gamma_{78} + \Gamma_{9\underline{10}} \right) \, \chi} e^{ \frac 12  \Gamma_{7\underline{10}} \xi_1 } e^{- \frac 12 \Gamma_{89} \xi_2} \epsilon_0 \equiv M \epsilon_0 \,,
		\end{align}
		where $\gamma$ denotes the four-product $\Gamma_{789\underline{10}}$. 
		\vspace{.1cm}

\section{Proof of the general giant-like solution in \eqref{newgensolution}}\label{AppA}

\noindent
In this appendix, we give the proof of our result in \eqref{newgensolution}. The M5 brane solution that we seek has a worldvolume 3-form field strength turned on. The value of the field strength will remain proportional to the size of the $S^3$ sphere wrapped by the world volume \eqref{sizedependnth}. 
The $\kappa$-symmetry equation is given by 
\begin{align}
		\Gamma_{\kappa} \epsilon \, = \, \sqrt{- \text{det} g} \, \epsilon
\end{align}
%
%
%
with $\Gamma_{\kappa}$ equal to 
\begin{align}
		\Gamma_{\kappa} \, = \, \gamma_{\tau 12345} \, + \, \gamma_{\tau 12} \, h_{\chi \xi_1 \xi_2} \, - \, \gamma_{345} \, h_{\tau 12} \,.
\end{align}
%
The world volume coordinates $\sigma_3, \, \sigma_4, \, \sigma_5$ are always going to be $\chi$, $\xi_1$, and $\xi_2$, respectively, which are on $S^3$ $\subset S^4$ of $AdS_7 \times S^4$. Therefore, the factors $h_{\chi \xi_1 \xi_2}$ and $\gamma_{345}$ are fixed constants here, proportional to the volume of the $S^3$. And $\gamma_{345}$ is simply equal to $ \Omega_3 \, \Gamma_{89\underline{10}}$. So $\Gamma_{\kappa}$ can also be written as
\begin{align}\label{GammakinApp}
\Gamma_{\kappa} \, = \, \Omega_3 \, \left[ \gamma_{\tau12} \Gamma_{89\underline{10}} \, + \, a \, \gamma_{\tau12} - a \, \Gamma_{89\underline{10}} \left( \star_g \Omega_3 \right)_{\tau12} \right]
\end{align}
where $a$ is the same parameter we saw in section \ref{giant-like} which is equal to $\frac{1 - \sin \theta}{\cos \theta}$. The worldvolume coordinates $\tau$, $\sigma_1$ and $\sigma_2$ are chosen from the $AdS_7$ part. And we can again rewrite \eqref{GammakinApp} as
\begin{align}\label{GammakinApp2}
\Gamma_{\kappa} \, =& \, \Omega_3 \, \left[ \mathfrak{e}^a_{\tau}  \mathfrak{e}^b_{1} \mathfrak{e}^c_{2} \, \Gamma_{abc} \Gamma_{89\underline{10}} \, + \, a \, \mathfrak{e}^a_{\tau}  \mathfrak{e}^b_{1} \mathfrak{e}^c_{2} \, \Gamma_{abc} - a \, \Gamma_{89\underline{10}} \left( \star_g \Omega_3 \right)_{\tau12} \right] \cr
=& \, \Omega_3 \, \mathfrak{e}^a_{\tau}  \mathfrak{e}^b_{1} \mathfrak{e}^c_{2} \, \left[  \Gamma_{abc} \Gamma_{89\underline{10}}  \, + \, a \,  \Gamma_{abc} \right] \, + \, a \,\sqrt{- \text{det} \, g} \, \Gamma_{89\underline{10}} 
\end{align}
(here we have introduced the notation $\mathfrak{e}^a_i$ for pull-backs of the 11d frame vielbein as $\mathfrak{e}^a_i \, \equiv \, e^a_{\mu} \partial_i X^{\mu}$).\\

\noindent
In the next step of this proof, we will show what the factor $\mathfrak{e}^a_{\tau} \mathfrak{e}^b_{1} \mathfrak{e}^c_{2} \, \Gamma_{abc}$ yields in the l.h.s. of the $\kappa$-symmetry equation. For the four independent projection conditions from \eqref{onesixteenthBPS} 
\begin{align}
		\label{116BPSprojections}
		\Gamma_{14} \, \epsilon_0 \, = \, \Gamma_{25} \, \epsilon_0 \, = \, \Gamma_{36} \, \epsilon_0 \, = \, - \Gamma_0 \, \epsilon_0 \, = \,  i \, \epsilon_0 \, 
\end{align}
the killing spinor reduces to this simplified form 
 \begin{align}\label{KSreducedgen}
    \epsilon \, = \, e^{ - \frac i2 \left( \phi_0 \, + \,3 \phi_1  \right) } e^{\frac 12 \gamma \Gamma_7 \theta} e^{- \frac 12 \left( \xi_1 \, + \, \xi_2 \right) \Gamma_{89}} \epsilon_0 \, = \,M \epsilon_0 \,.
 \end{align}
Now we expand the term $\mathfrak{e}^a_{\tau} \mathfrak{e}^b_{1} \mathfrak{e}^c_{2} \, \Gamma_{abc}$ in the $\kappa$-symmetry equation
\begin{align}
		\label{gengkappaM2}
		\Gamma_{abc} \, \mathfrak{e}^a \wedge \mathfrak{e}^b \wedge \mathfrak{e}^c \, \epsilon \, =& \, \Big[ \mathfrak{e}^{012} \, \Gamma_{012} \, + \, \mathfrak{e}^{013} \, \Gamma_{013} \, + \,\mathfrak{e}^{014} \, \Gamma_{014} \, + \,\mathfrak{e}^{015} \, \Gamma_{015} \, + \,\mathfrak{e}^{016} \, \Gamma_{016} \, + \,\mathfrak{e}^{023} \, \Gamma_{023} \, + \,\mathfrak{e}^{024} \, \Gamma_{024}  \cr
		& \, + \, \ldots \, + \, \ldots \, \ldots \, +  \,\mathfrak{e}^{346} \, \Gamma_{346} \, + \,\mathfrak{e}^{356} \, \Gamma_{356} \, + \,\mathfrak{e}^{456} \, \Gamma_{456} \Big] \, \epsilon \,.
\end{align}
We find 35 component terms. After using the 4 independent projections in \eqref{gengkappaM2}, the terms inside square bracket give
\begin{align}\label{etau12Gammaabc}
		& \Gamma_{12} \left( - i \, \mathfrak{e}^{012} \, + \,  \mathfrak{e}^{015} \, - \,  \mathfrak{e}^{024} \, + \, i \, \mathfrak{e}^{045}   \right) \, + \, \Gamma_{13} \left( - i \, \mathfrak{e}^{013} \, + \,  \mathfrak{e}^{016} \, - \,  \mathfrak{e}^{034} \, + \, i \, \mathfrak{e}^{046}   \right) \cr
		+ \, & \Gamma_{23} \left( - i \, \mathfrak{e}^{023} \, + \,  \mathfrak{e}^{026} \, - \,  \mathfrak{e}^{035} \, + \, i \, \mathfrak{e}^{056}   \right)  \, + \, \Gamma_2 \left( - i \, \mathfrak{e}^{124} \, - \, \mathfrak{e}^{145} \, + \, i \, \mathfrak{e}^{236} \, + \, \mathfrak{e}^{356} \right) \cr 
		+ \, & \Gamma_1 \left(  i \, \mathfrak{e}^{125} \, + \,i \, \mathfrak{e}^{136} \, + \, \mathfrak{e}^{245} \, + \, \mathfrak{e}^{346} \right) \, + \, \Gamma_3 \left( - i \, \mathfrak{e}^{134} \, - \, \mathfrak{e}^{146} \, - \, i \, \mathfrak{e}^{235} \, - \, \mathfrak{e}^{256} \right) \cr 
		+ \, & \Gamma_{123} \left( \mathfrak{e}^{123} \, + \, i \, \mathfrak{e}^{126} \, - \,i\, \mathfrak{e}^{135} \, - \, \mathfrak{e}^{156} \, + \, i \, \mathfrak{e}^{234} \, + \, \mathfrak{e}^{246} \, - \, \mathfrak{e}^{345} \, - \, i \, \mathfrak{e}^{456}\right) \, + \, \mathfrak{e}^{014} \, + \, \mathfrak{e}^{025} \, + \, \mathfrak{e}^{036} \,.\cr
\end{align}
We put the coefficients of every $\Gamma$ factor to zero in the above equation. And with the help of the following complex 1-form definitions 
\begin{align}
\label{complex-1form1}
		\mathbf{E}^1 \, = \, \mathfrak{e}^1  \, - \, i \,\mathfrak{e}^4 \qquad \mathbf{E}^2 \, = \, \mathfrak{e}^2  \, - \, i \,\mathfrak{e}^5 \qquad \mathbf{E}^3 \, = \, \mathfrak{e}^3  \, - \, i \,\mathfrak{e}^6 \,,
\end{align}
we write down every constraint in the parenthesis brackets in \eqref{etau12Gammaabc} in these compact forms
\begin{align}
		\label{3-formconstraints}
		\mathfrak{e}^0 \wedge \overline{\mathbf{E}}^a \wedge \overline{\mathbf{E}}^b \, = \, 0 \qquad \quad \overline{\mathbf{E}}^a \wedge \omega \, = \, 0 \qquad \quad \overline{\mathbf{E}}^1 \wedge \overline{\mathbf{E}}^2 \wedge \overline{\mathbf{E}}^3 \, = \, 0 \,.
\end{align}
These are 3 + 3 + 1 = 7 equations and here $\omega$ is a real 2-form defined as $$\omega \equiv \mathfrak{e}^{14} \, + \,  \mathfrak{e}^{25} \, + \,  \mathfrak{e}^{36}\,.$$
This exercise with the BPS constraints \eqref{3-formconstraints} suggests to us that the volume form $\mathfrak{e}^a_{\tau}\mathfrak{e}^b_{1}\mathfrak{e}^c_{2} \, \varepsilon_{abc}$ in the 3-directions $\tau$, $\sigma_1$, $\sigma_2$ (on the M5 world volume) is equal to
$$\mathfrak{e}^{014} \, + \, \mathfrak{e}^{025} \, + \, \mathfrak{e}^{036}\,.$$

For the $\frac1{16}$ BPS embedding solution that we are after, has the $S^4$ polar coordinate $\theta$ fixed to an arbitrary value. And the coordinates of $S^3$ wrapped by the world volume are identified with $\chi$, $\xi_1$, $\xi_2$ of $S^4$. So we just need 4 real conditions to describe the embedding of this general M5 brane solution. Therefore, we consider two arbitrary complex functional conditions
\begin{align}
F^{(I)} \left( \phi_0, \rho, \alpha, \beta, \phi_1, \phi_2, \phi_3\right) \, = \, 0 \,.
\end{align}
This leads to the differential constraints
\begin{align}
P \left[ F^{(I)}_{\rho} \, d\rho + F^{(I)}_{\alpha} \, d\alpha  + F^{(I)}_{\beta}  \, d\beta + \sum_{i=0}^3 F^{(I)}_{\phi_i} \, d\phi_i \right] = 0
\end{align}
where $P$ denotes pullback onto the world volume. We rewrite this in terms of the complex one-forms defined in \eqref{complex-1form1} using the frame vielbeins in \eqref{adsframe} and \eqref{S4frame} (given in the appendix)
\begin{align} \label{onedifferentialconstraint}
& \mathbf{E}^1 \left( F^{(I)}_{\rho} - i  \sum_{i=1,2,3} F^{(I)}_{\phi_i} \, \coth \rho - i \, F^{(I)}_{\phi_0} \tanh \rho \right)  
+  \overline{\mathbf{E}^1} \left( F^{(I)}_{\rho} + i  \sum_{i=1,2,3} F^{(I)}_{\phi_i} \, \coth \rho + i \, F^{(I)}_{\phi_0} \tanh \rho \right) \cr
+ \, &\sinh^{-1} \rho \left[ \mathbf{E}^2 \left(  F^{(I)}_{\alpha} - i \sum_{i=2,3} F^{(I)}_{\phi_i} \cot \alpha + i \, F^{(I)}_{\phi_1} \tan \alpha \right) + \overline{\mathbf{E}^2} \left(  F^{(I)}_{\alpha}  + i \sum_{i=2,3} F^{(I)}_{\phi_i} \cot \alpha - i \, F^{(I)}_{\phi_1} \tan \alpha \right) \right] \cr
+ \, & \, \csc \alpha  \sinh^{-1} \rho \bigg[ \mathbf{E}^3  \left( F^{(I)}_{\beta} - i \, F^{(I)}_{\phi_3} \cot \beta + i \, F^{(I)}_{\phi_2} \tan \beta \right) + \overline{\mathbf{E}^3} \left( F^{(I)}_{\beta} + i \, F^{(I)}_{\phi_3} \cot \beta - i \, F^{(I)}_{\phi_2} \tan \beta \right) \bigg] \cr 
& \hspace{4cm}  +   \, \mathfrak{e}^0 \,  \sum_{i=0}^3 F^{(I)}_{\phi_i}  \,  = \, 0 \,.
\end{align}
As the next step, we solve for $\mathfrak{e}^0$ from above \eqref{onedifferentialconstraint} and substitute in the complex conjugate of one of the equations in \eqref{3-formconstraints} which is $$\mathfrak{e}^0 \wedge \mathbf{E}^1 \wedge \mathbf{E}^a \, = \, 0 \,.$$ On account of the remaining 3-form constraints which are
\begin{align}
 \mathfrak{e}^0 \wedge \mathbf{E}^2 \wedge \mathbf{E}^3 \, = \, 0 \qquad \quad \mathbf{E}^a \wedge \omega \, = \, 0 \qquad \quad \mathbf{E}^1 \wedge \mathbf{E}^2 \wedge \mathbf{E}^3 \, = \, 0
\end{align}
the l.h.s. $\mathfrak{e}^0 \wedge \mathbf{E}^1 \wedge \mathbf{E}^a$ vanishes, if we also consider the following differential equations which are the coefficients of $\overline{\mathbf{E}}^1$, $\overline{\mathbf{E}}^2$ and $\overline{\mathbf{E}}^3$ in \eqref{onedifferentialconstraint}
\begin{align}
\label{differentialcontraintM2gen}
F^{(I)}_{\rho} + i  \sum_{i=1,2,3} F^{(I)}_{\phi_i} \, \coth \rho + i \, F^{(I)}_{\phi_0} \tanh \rho &= 0 \cr
F^{(I)}_{\alpha}  + i \sum_{i=2,3} F^{(I)}_{\phi_i} \cot \alpha - i \, F^{(I)}_{\phi_1} \tan \alpha  &= 0 \cr
F^{(I)}_{\beta} + i \, F^{(I)}_{\phi_3} \cot \beta - i \, F^{(I)}_{\phi_2} \tan \beta &= 0 \,.
\end{align}	
On solving these in \eqref{differentialcontraintM2gen}, we find that the functional conditions $F^{(I)}$ have holomorphic dependence on $AdS_7$ complex coordinates (defined in \eqref{complexAdS7})
\begin{align}
\label{generalsolutions1by16}
F^{(I)} ( \Phi_0 \,, \Phi_1\,, \Phi_2 \,, \Phi_3) = 0 \,
\end{align}
 for $I$ equal to $1$ and $2$.\\

\noindent
In addition to \eqref{differentialcontraintM2gen} the functional constraints $F^I$ also satisfy the scaling condition
\begin{align}
   \sum_{i=0}^3 \partial_{\phi_i} F^{(I)} ( \Phi_0 \,, \Phi_1\,, \Phi_2 \,, \Phi_3) = 0 \,.
\end{align}
This can be seen when we pick a 1-form $\mathbf{E}^a$ and solve for it from \eqref{onedifferentialconstraint} in terms of other 1-forms and substitute in the 3-form constraint 
\begin{align}
    \mathbf{E}^a \wedge \omega \, = \, 0 \,.
\end{align}

\vspace{.2cm}

 \noindent
 Now we go back to the expression for $\Gamma_{\kappa}$ in \eqref{GammakinApp2} and consider its action on the reduced killing spinor in \eqref{KSreducedgen}. After taking into account the BPS constraint of \eqref{3-formconstraints} and the projection conditions \eqref{116BPSprojections} on $\epsilon_0$, for the l.h.s. of $\kappa$-symmetry equation we get 
 \begin{align}
    \Gamma_{\kappa} \, \epsilon \, =& \, \sqrt{- \text{det} \, g} \, M \, \left[ e^{-  \gamma \Gamma_7 \theta} \left( - \Gamma_{89\underline{10}} \, + \, a \, \right) \, + \, a \, \Gamma_{89\underline{10}} \right] \, \epsilon_0 \, \cr
    =& \, \sqrt{- \text{det} \, g} \, M \, \left[ - \cos \theta \, \Gamma_{89\underline{10}} \, + \, \sin \theta \,+ \, a \left( \cos \theta \, + \, \Gamma_{89\underline{10}} \, \sin \theta  \right) \, + \, a \, \Gamma_{89\underline{10}} \right] \, \epsilon_0 \,.
 \end{align}
 For $a = \frac{1 - \sin \theta}{\cos \theta}$ this is equal to $\epsilon$ and we confirm the supersymmetry of the solution in \eqref{newgensolution}.
\noindent

%
%

  \providecommand{\href}[2]{#2}\begingroup\raggedright\endgroup

\end{document}